\documentclass[letterpaper]{article} 
\usepackage{aaai2027}  
\usepackage[hyphens]{url}  
\usepackage{graphicx} 
\usepackage{natbib}  
\usepackage{caption} 
\usepackage{amsmath}
\usepackage{amssymb}
\usepackage{array}
\usepackage{booktabs}

\newcommand{\rqanswer}[2]{\par\noindent\textbf{#1.} #2\par}

\title{Trajectory-Level Security Debt in LLM Coding Agents}

\author{
    Prateek Rajput\textsuperscript{\rm 1},
    Abdoul Kader Kabore\textsuperscript{\rm 1},
    Yewei Song\textsuperscript{\rm 1},
    M\'{e}lissa Tessa\textsuperscript{\rm 1},\\
    Tailia Malloy\textsuperscript{\rm 1},
    Jacques Klein\textsuperscript{\rm 1},
    Tegawend\'{e} F. Bissyand\'{e}\textsuperscript{\rm 1}
}
\affiliations{
    \textsuperscript{\rm 1}University of Luxembourg, Esch-sur-Alzette, Luxembourg\\
    \{prateek.rajput, abdoulkader.kabore, yewei.song, melissa.tessa,\\
    tailia.malloy, jacques.klein, tegawende.bissyande\}@uni.lu
}

\begin{document}

\maketitle

\begin{abstract}
LLM coding agents can traverse hundreds of intermediate code states before submitting a solution. Evaluating only the final artifact leaves the evolution of security findings unmeasured. We introduce the Security Debt Line Integral (SDLI), which accumulates static-analysis risk when an agent reaches a new best test pass ratio. We instantiate it with four static application security testing (SAST) tools and study artifacts from 830 passing SWE-bench runs, 712 ProgramBench final workspaces, and 13 public MirrorCode trajectories. The two large populations use the final-state special case of SDLI. Two-tool Common Weakness Enumeration (CWE) class agreement occurs in 3.9\% of SWE-bench runs and 26.2\% of the 80 ProgramBench runs passing at least 90\% of official tests. These are scanner findings, not validated vulnerability rates. Excluding three advisory-heavy classes reduces the latter rate to 6.2\%. Same-task runs differ in their measured scores, while one reconstructed ProgramBench run exposes persistent findings from its first implementation write. A repair case study reduces the scanner signal while preserving tested behavior, but also reveals sensitivity to equivalent API rewrites. SDLI offers a way to study progress and security findings together. Its value for steering agents and confirming exploitable vulnerabilities remains to be established.

\end{abstract}

\section{Introduction}
\label{sec:intro}

Code reuse can spread the consequences of an insecure design choice across a repository~\citep{lopes2017dejavu,mojica2013large,prana2021out} and through dependencies~\citep{zerouali2022impact,zimmermann2019small}. LLM-generated code is susceptible to such choices. Copilot suggestions were vulnerable in roughly 40\% of the tested CWE scenarios~\citep{pearce2025asleep}, and subsequent work examines security in developer assistance and iterative generation~\citep{sandoval2023lost,shukla2025security,schreiber2025security,tihanyi2023formai,wu2023deceptprompt}. As coding agents reuse their own earlier output, security-relevant decisions can persist through many later changes.

LLM coding agents now write repositories over hundreds or thousands of turns. MirrorCode traces span roughly 12{,}000 messages~\citep{mirrorcode2025}, SWE-EVO tasks touch 21 files on average~\citep{le2025swe}, and NL2Repo-Bench requires at least 181 interaction turns to produce an installable library~\citep{ding2025nl2repo}. Recorded actions offer checkpoints that can be scanned before submission. Scanning only the submitted artifact leaves these intermediate states unobserved. Our focus is the coupling of checkpoint findings with measured functional progress.

\begin{figure*}[t]
\centering
\includegraphics[width=\textwidth]{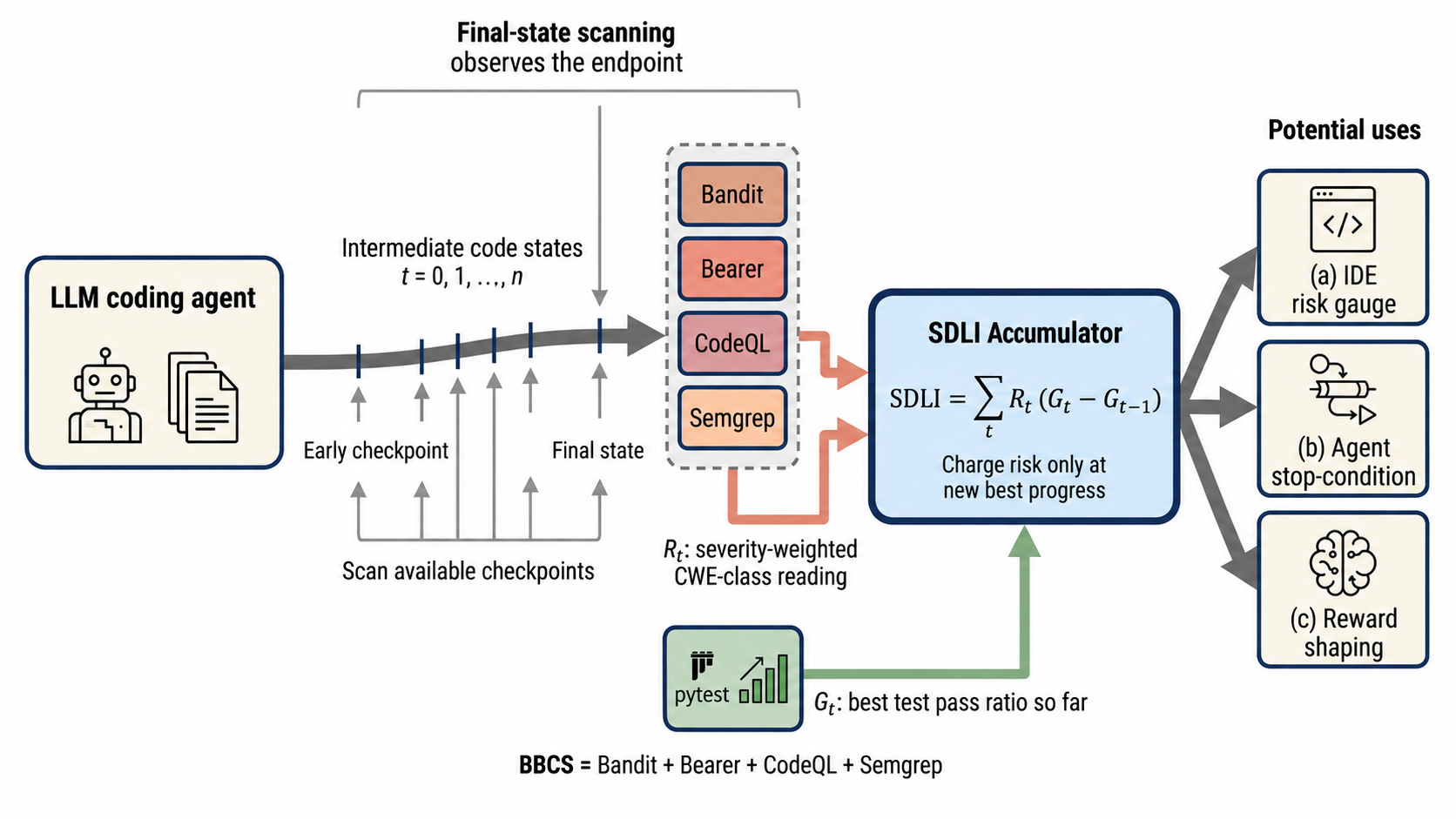}
\caption{The SDLI pipeline and potential uses. At available checkpoints, SAST readings are paired with increases in historical-best progress. The endpoint special case uses one final scan. The applications shown are proposals, not evaluated deployments.}
\label{fig:sdli_pipeline}
\end{figure*}

This paper takes a first step toward closing that gap. We formalize trajectory-level security measurement as the Security Debt Line Integral (SDLI), which charges a per-checkpoint risk reading against the functional progress banked at that checkpoint. A risky checkpoint incurs no charge if it does not improve historical-best progress. Charges already accumulated remain even if the code is later removed. We instantiate the risk reading via BBCS, a four-tool SAST oracle named after its constituent tools Bandit, Bearer, CodeQL, and Semgrep, identified as Pareto-optimal for Python by \citet{liu2026sastpython}. Figure~\ref{fig:sdli_pipeline} shows the pipeline.

We evaluate SDLI on three populations and, throughout the paper, restrict headline analysis to runs whose code actually works. This filter asks whether useful solutions also carry security findings. Findings in unsuccessful runs remain relevant, but answer a different question. The two large populations are 830 passing short-horizon SWE-bench trajectories and, from 712 scanned Python ProgramBench solutions produced by six frontier model configurations, the 80 mostly working ones that pass at least 90\% of the official test suite. The 13 public MirrorCode runs, of which 3 reach a recorded pass ratio of 1.0, expose up to 146 scannable checkpoints per run and serve as a focused case study of how debt enters and persists over weeks-scale trajectories. Our goal is to quantify how much security debt agent trajectories accumulate, to understand when and how it enters, and to test whether it can be repaired once a scanner points at it. For the last question we replay a real debt-carrying trajectory at the step where the scanner findings first appeared and hand the scanner's findings back to a model.

\section{Background and Research Questions}
\label{sec:background}

\subsection{Long-Horizon Agent Tasks}

Benchmarks have expanded from isolated functions in HumanEval~\citep{chen2021evaluating} to repository-level issue resolution~\citep{jimenez2023swe,yang2024swe} and longer tasks in SWE-EVO, NL2Repo-Bench, MirrorCode, and ProgramBench~\citep{le2025swe,ding2025nl2repo,mirrorcode2025,yang2026programbench}. Similar workflows are entering mainstream development~\citep{ray2025review,lipsanen2026shift}. Maintaining human oversight across long runs remains difficult~\citep{takerngsaksiri2025human}, and multi-agent pipelines introduce further failure modes~\citep{cemri2025multi}. Checkpoint monitoring offers a way to inspect security findings while the implementation is still evolving.

\subsection{Why Python and Why BBCS}
\label{sec:why_python}

We focus on Python because it has established SAST tools and public agent artifacts. Tool choice determines which weakness classes are visible. We adopt Bandit~\citep{bandit}, Bearer~\citep{bearer}, CodeQL~\citep{codeql}, and Semgrep~\citep{semgrep}, the BBCS combination studied by \citet{liu2026sastpython}. Their empirical evaluation motivates combining tools, but does not validate precision on our agent-generated artifacts.

\subsection{Related Work}
\citet{shukla2025security} study security degradation during iterative LLM code refinement. SDLI adds an explicit coupling between scanner readings and functional progress across recorded agent states. The broader study of when defects enter a history predates coding agents. SZZ links fixes to earlier changes~\citep{sliwerski2005changes}, and vulnerability-contributing-commit studies examine security-relevant changes and prioritize audits~\citep{meneely2013patch,perl2015vccfinder}. Our class-level measurement does not provide that commit-level attribution.

SecurityEval~\citep{siddiq2022securityeval} and CyberSecEval~\citep{bhatt2023cyberseceval} evaluate insecure code generation. BaxBench evaluates both backend functionality and security with executable exploits~\citep{vero2025baxbench}. SDLI instead summarizes the timing of SAST findings relative to progress. These approaches are complementary, and exploit-based validation would strengthen the interpretation of our scores.

\subsection{Research Questions}

SDLI combines scanner readings with progress, with trajectory detail limited by the available checkpoints. We use it to ask four questions.

\begin{itemize}
  \item[\textbf{RQ1}] To what extent do functionally correct agent trajectories accumulate security debt across short- and long-horizon tasks?
  \item[\textbf{RQ2}] Do different runs solving the same task produce meaningfully different SDLI?
  \item[\textbf{RQ3}] How does security debt enter a trajectory? Does it build up slowly or arrive suddenly at a single step, and do agents ever remove it on their own?
  \item[\textbf{RQ4}] Once the debt is pointed out, can it be repaired without losing the functional progress the agent has already made?
\end{itemize}

\section{The SDLI Metric}
\label{sec:method}

Let $g_t\in[0,1]$ be the observed test pass ratio at checkpoint $t$, and let $R_t\geq0$ be its BBCS risk reading. We define banked progress as the historical maximum, $G_t=\max_{1\leq s\leq t}g_s$ for $t\geq1$, with $G_0=0$ as an accounting baseline. The Security Debt Line Integral is
\begin{equation}
\label{eq:sdli}
\mathrm{SDLI}=\sum_{t=1}^{n}R_t\,(G_t-G_{t-1}).
\end{equation}
This definition makes the progress envelope used by our implementation explicit. For observed progress $0,0.8,0.4,0.8$ at constant risk 1, SDLI is 0.8. Recovering previously achieved progress adds no charge.

SDLI measures cumulative progress-weighted scanner exposure. It is not the outstanding risk in the current state, which is $R_t$, or a time-weighted exposure measure. Risk introduced after progress stops increasing contributes zero. Removing a finding lowers the current reading but does not refund prior charges. A lower-risk path has lower SDLI only if the reduction occurs at checkpoints that add progress. Monitoring should therefore retain the $R_t$ trace and final reading alongside SDLI.

The risk reading itself aggregates the raw scanner output at checkpoint $t$. Let $\mathcal{C}_t$ be the set of CWE classes that at least one BBCS tool reports at that checkpoint, and for a class $c \in \mathcal{C}_t$ let $\mathcal{T}_t(c)$ be the set of tools that flagged it. Each tool $k \in \mathcal{T}_t(c)$ rates the class with a severity level $\mathrm{sev}_k(c) \in \{\text{LOW}, \text{MEDIUM}, \text{HIGH}, \text{CRITICAL}\}$, and the weighting function $\mathrm{sev\_w}$ turns the level into a number, $3.0$ for HIGH or CRITICAL, $1.0$ for MEDIUM, and $0.25$ for LOW. At agreement threshold $a \in \{1,2\}$ the reading counts every class flagged by at least $a$ tools exactly once, at the weight of the most severe rating any tool gave it,
\begin{equation}
R_t^{(a)} \;=\; \sum_{\substack{c \in \mathcal{C}_t \\ |\mathcal{T}_t(c)| \geq a}} \mathrm{sev\_w}\!\bigl(\max_{k \in \mathcal{T}_t(c)} \mathrm{sev}_k(c)\bigr).
\end{equation}
Threshold $a{=}1$ gives BBCS-union, where one tool is enough, and $a{=}2$ gives BBCS-agree2, where at least two tools report the same normalized class. We use agree2 as a stricter reporting threshold, without assuming independent errors or validated precision. Agreement can involve different locations or data flows. Counting each class once also discards the number of affected sites. Throughout the paper, security debt refers to this scanner-based construct, not a count of confirmed vulnerabilities.

\section{Benchmarks and Measurement Setup}
\label{sec:setup}

We evaluate SDLI on three benchmarks spanning short- and long-horizon agent tasks. Table~\ref{tab:datasets} summarizes the three populations.

\textbf{SWE-bench}~\citep{jimenez2023swe} trajectories are drawn from the publicly released SWE-Agent runs\footnote{\url{https://huggingface.co/datasets/nebius/SWE-agent-trajectories}} across three LLaMA models (8B, 70B, 405B), covering 830 functionally correct runs on 206 unique instances.

\textbf{ProgramBench}~\citep{yang2026programbench} asks agents to reimplement whole CLI programs from binaries and documentation, and publishes frontier-model submissions at scale. We scan 712 Python final workspaces from six public submissions\footnote{\url{https://github.com/ProgramBench/submissions}, with heavy artifacts hosted on Hugging Face.} (Claude Opus 4.7, Claude Sonnet 4.6, Gemini 3.1 Pro, and GPT-5.5 at three reasoning-effort settings), each paired with a full agent trajectory and an official evaluation pass rate. Unlike SWE-bench issue fixes, no scanned ProgramBench run passes 100\% of the official hidden tests (the best reaches 99.7\%). We therefore take the 80 mostly working runs with $G_T \geq 0.90$ as the primary population and use the 24 near-solved runs with $G_T \geq 0.95$ as a stricter subset.

\textbf{MirrorCode}~\citep{mirrorcode2025} provides all 13 public Python trajectories\footnote{\url{https://epoch.ai/blog/mirrorcode-preliminary-results}} generated by Claude Opus (4.6 for the gotree run reaching recorded progress 1.0, an earlier Opus version for the remaining 12), spanning CLI reimplementation tasks up to roughly 604M tokens and 146 checkpoints per run. Three runs reach a recorded pass ratio of 1.0 (cal, choose, and the gotree run by Opus 4.6). We use this recorded-progress criterion for the working subset, rather than claim independently verified full-suite success.

\begin{table*}[t]
\centering
\small
\setlength{\tabcolsep}{2.2pt}
\begin{tabular}{llll}
\toprule
                        & \textbf{SWE-bench}    & \textbf{ProgramBench}   & \textbf{MirrorCode} \\
\midrule
Runs scanned            & 830                   & 712           & 13 \\
Working subset          & 830 (passing)         & 80 ($G_T{\geq}0.90$)    & 3 (recorded $G_T{=}1$) \\
Unique tasks            & 206                   & 187                     & 5 \\
Models                  & LLaMA 8B/70B/405B     & 6 frontier configs      & Claude Opus (2) \\
Agent scaffold          & SWE-Agent             & mini-SWE-agent          & Epoch AI harness \\
Agent turns             & 3 to 15               & 11 to 986               & 6 to 6{,}084 \\
BBCS checkpoints        & 1 (final extract)        & 1 (final tree)          & up to 146 \\
Horizon                 & Short                 & Long           & Long \\
\bottomrule
\end{tabular}
\caption{Benchmark and model summary.}
\label{tab:datasets}
\end{table*}

\textbf{Scan scope and progress.} We scan the available code snapshot, not just changed lines. ProgramBench uses final workspaces. SWE-bench uses final code states extracted from file content visible in the agent transcript, so these are not guaranteed to be complete repositories. Although its parser records two to five checkpoints per run, the four-tool results reported here scan only the final extracted state. Both large populations therefore use the endpoint special case $R_TG_T$ with a zero accounting baseline. ProgramBench progress comes from the official hidden-test evaluation. SWE-bench uses the parser's final progress envelope, which can include agent-run test subsets. The solved-run selection is separate from that progress proxy.

MirrorCode progress comes from test results recorded in its public transcripts, carried into a historical-best envelope. The 463 available code snapshots provide trajectory measurements, but changing test coverage limits comparability with an independently run fixed suite. Bandit, Bearer, and Semgrep scan each snapshot. CodeQL scans selected checkpoints, with interpolation between them. It reports zero risk throughout the available MirrorCode scans, so interpolation contributes no signal here. The gowsdl replay below reconstructs per-action risk, not an official per-action progress curve.

\textbf{Endpoint interpretation.} The endpoint score is not a lower bound on full SDLI. If all progress is achieved at risk 0 and risk appears only afterward, full SDLI is 0 while $R_TG_T>0$. Conversely, risk charged during progress may later disappear, giving $R_TG_T=0$ with positive SDLI. Only the set of findings observed across checkpoints necessarily contains the final set, under the same scan configuration.

\textbf{Scanner configuration.} Bandit skips B101, B110, B112, and B401 through B413. Bearer uses its SAST scanner with all severity levels. Semgrep uses \texttt{p/python}, \texttt{p/security-audit}, and \texttt{p/owasp-top-ten}. CodeQL uses the Python security-extended suite, with a configured fallback to query pack 1.3.0. The artifact records the remaining query filters. CWE identifiers are normalized from tool metadata, with an additional Bandit rule mapping. The mapping groups related identifiers, including CWE-78 and CWE-88 into CWE-77, and CWE-23 through CWE-41 into CWE-22. These are analysis groupings, not evidence that two findings describe the same defect. The full mapping and scan commands are in the artifact.

Bandit, Bearer, and Semgrep typically take under 10 seconds each per snapshot in our runs. CodeQL takes roughly two to five minutes. These are observed scan times, not an end-to-end overhead study. We have not measured sensitivity to checkpoint frequency, weights, or ruleset revisions. Remote registry configurations were not all pinned to immutable revisions, which limits exact rescanning.

\section{Results at Scale}
\label{sec:results}

\begin{figure*}[t]
\centering
\includegraphics[width=0.98\textwidth]{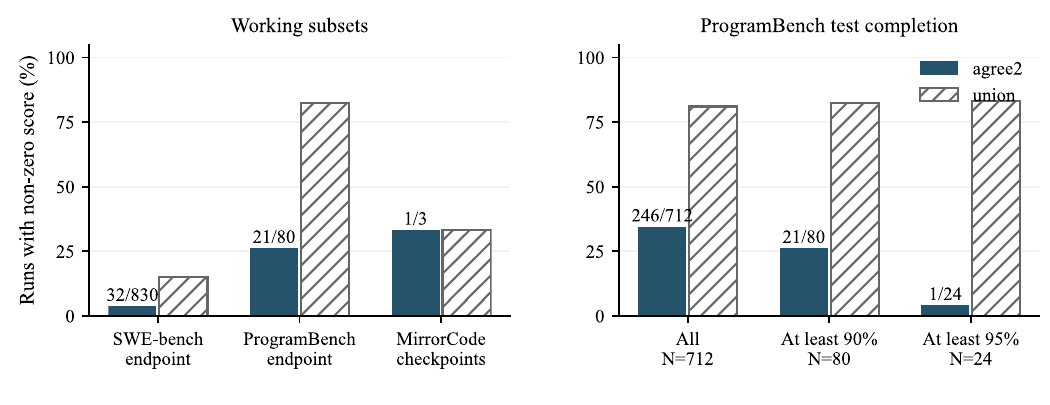}
\caption{Scanner-positive proportions in distinct benchmark subsets (left), and nested ProgramBench completion tiers (right). Counts are shown for agree2. SWE-bench and ProgramBench use endpoint scores, while MirrorCode uses recorded checkpoints. The groups differ in tasks, progress measures, and sample sizes.}
\label{fig:results_overview}
\end{figure*}

Our two large populations answer RQ1 and RQ2. Table~\ref{tab:agg_results} aggregates both working populations, and Figure~\ref{fig:results_overview} summarizes the distinct populations and shows how ProgramBench debt behaves as functional completeness rises. We report nominal 95\% Wilson intervals to show count uncertainty. They assume independent runs and are descriptive only, since runs share tasks and models. For MirrorCode, 1/3 gives an interval from 6.1\% to 79.2\%, which emphasizes the limited sample.

\begin{table*}[t]
\centering
\small
\setlength{\tabcolsep}{3.4pt}
\begin{tabular}{lrrrr}
\toprule
 & \multicolumn{2}{c}{\textbf{SWE-bench}} & \multicolumn{2}{c}{\textbf{ProgramBench}} \\
\cmidrule(lr){2-3}\cmidrule(lr){4-5}
\textbf{Metric} & \textbf{agree2} & \textbf{union} & \textbf{agree2} & \textbf{union} \\
\midrule
Runs analyzed          & 830           & 830           & 80           & 80 \\
Non-zero SDLI          & 32 (3.9\%)    & 124 (14.9\%)  & 21 (26.2\%)  & 66 (82.5\%) \\
95\% interval (\%)     & 2.7 to 5.4 & 12.7 to 17.5 & 17.9 to 36.8 & 72.7 to 89.3 \\
Mean SDLI (non-zero)   & 2.82          & 2.88          & 3.28         & 4.12 \\
Max SDLI               & 3.00          & 6.21          & 10.85        & 12.66 \\
\bottomrule
\end{tabular}
\caption{Endpoint SDLI for passing SWE-bench runs and ProgramBench runs passing at least 90\% of official tests. Both use final-state BBCS scans.}
\label{tab:agg_results}
\end{table*}

\subsection{Short-Horizon Patches on SWE-bench}

On SWE-bench, 3.9\% of the selected test-passing runs contain findings with two-tool class agreement, and the recall-favoring union reading raises the rate to 14.9\%. These are issue-fixing tasks in existing projects. We do not establish whether the flagged code was introduced by the agent. The dominant two-tool-reported classes are CWE-502 (insecure deserialization, 18 runs) and CWE-327 (weak crypto, 10 runs). The union adds a single-tool tail led by CWE-22 path handling (46 runs) and CWE-94 code injection (23 runs). The observed findings show that test success can coexist with a non-zero scanner reading.

\begin{table}[t]
\centering
\small
\setlength{\tabcolsep}{2.6pt}
\begin{tabular}{lrcccc}
\toprule
 & & \multicolumn{2}{c}{agree2} & \multicolumn{2}{c}{union} \\
\cmidrule(lr){3-4}\cmidrule(lr){5-6}
Instance & Runs & Min & Max & Min & Max \\
\midrule
hatch-422             & 9  & 0.00 & 3.00 & 0.00 & 3.00 \\
sphinx-gallery-625    & 2  & 0.00 & 3.00 & 0.00 & 3.00 \\
msrest-for-python-208 & 9  & 3.00 & 3.00 & 3.00 & 6.00 \\
datalab-pandas-29     & 10 & 2.40 & 2.57 & 5.80 & 6.21 \\
vault-cli-111         & 9  & 0.00 & 0.00 & 0.00 & 3.25 \\
\bottomrule
\end{tabular}
\caption{SWE-bench instances with the largest same-task SDLI variation across runs, under both channels.}
\label{tab:swe_instances}
\end{table}

Table~\ref{tab:swe_instances} shows how much runs on the same task vary. On \texttt{pypa/hatch-422}, eight of nine runs bank an agree2 SDLI of 3.0 through the same pickle deserialization pattern (CWE-502) while one sibling has a zero score. All nine runs on \texttt{Azure/msrest-for-python-208} incur a two-tool-reported SDLI of 3.0, which is also consistent with a shared pre-existing finding or similar extracted code. On \texttt{peopledoc/vault-cli-111} the variation lives entirely in the union channel, where one run stays at zero while another accumulates 3.25.

\subsection{Whole-Program Synthesis on ProgramBench}

Among the 80 mostly working ProgramBench runs, 21 (26.2\%) have non-zero endpoint SDLI under agree2 and 66 (82.5\%) under union. The dominant agree2 class is CWE-77, reported by two tools in 18 of 80 runs. Bearer's path-handling rules contribute to the union signal on 45 runs. These classes include capability advisories that require contextual review, and they affect both channels.

To make this dependence explicit, we summarize the existing scans after excluding CWE-22, CWE-77, and CWE-939, the three advisory-heavy classes discussed in the case study. Agree2 remains positive in 5 of 80 runs (6.2\%), and union in 32 of 80 (40.0\%). This is a sensitivity summary, not a validated vulnerability estimate. Excluding an entire class can discard genuine defects as well as advisories. The 26.2\% rate should therefore not be read as confirmed insecurity or evidence of a horizon effect. The corpora also differ in task type, extraction coverage, models, and passing criterion.

\textbf{Debt versus functional completeness.} Figure~\ref{fig:results_overview} (right) tracks what happens as the functional bar rises, and the result complicates a simple ``long horizon means more debt'' story. Under the headline agree2 oracle, prevalence falls from 34.6\% across all 712 scanned finals to 26.2\% in the 80 runs above 90\% pass rate and to 4.2\% (a single run) among the 24 near-solved runs above 95\%, with mean agree2 SDLI falling to 0.12. The agree2 prevalence declines across these nested subsets. Their changing task and model composition prevents a causal interpretation. The union reading shows no such gradient (81.0\%, 82.5\%, 83.3\%), which is consistent with broad capability advisories remaining active across the subsets. The working tier's mean is pulled above the population's by the gowsdl case study below. Two readings are consistent with the agree2 gradient and our data cannot separate them. Lower scanner readings may accompany higher test performance, or the tier may reflect its composition, since 17 of its 24 runs come from one configuration (GPT-5.5 xhigh). What the gradient does not do is reach zero. One near-solved run still carries two-tool class agreement, and the MirrorCode gotree run reaching recorded progress 1.0 below carries the largest per-run score in its benchmark, so high recorded progress does not imply a zero scanner score.

On channel composition, Bandit and Bearer carry most of the union signal, Semgrep overlaps with some of these class-level findings, and CodeQL contributes findings on only 66 of the 712 finals and 6 of the 80 working runs. One practical note for anyone replaying this pipeline. Three of the four channels can silently report zero findings for purely environmental reasons. Semgrep and CodeQL skip symlinked directories, and Bearer skips any file that is gitignored in an enclosing repository, which includes scan workspaces stored under a repo's ignored output directory. Earlier passes of this corpus lost the Semgrep and CodeQL channels to the symlinks and the Bearer channel to the gitignore rule before per-file spot checks caught both, so we recommend planting a known-vulnerable canary file and verifying that every channel reports it before trusting any zero.

\subsection{Long-Horizon Traces on MirrorCode}

MirrorCode~\citep{mirrorcode2025} anchors the weeks-scale end. Of its 13 public Python trajectories, 3 reach a recorded pass ratio of 1.0, the cal and choose runs and the gotree run by Claude Opus 4.6. Two of the three have zero scores under both channels. The third, by far the largest of the three reimplementations at 7{,}810 lines, banks an agree2 SDLI of 3.000 and a union SDLI of 7.014, the highest per-run debt in its benchmark. This run combines high recorded progress with persistent scanner findings. Across the wider context population of all 13 runs, seven carry two-tool class agreement and ten accumulate non-zero BBCS-union SDLI, and six of ten gotree attempts across two Opus versions contain the same flagged XML parsing pattern, using \texttt{xml.etree.ElementTree} instead of \texttt{defusedxml} (CWE-611). Sibling trajectories on the same task have lower readings, although their functional outcomes differ. Four of the ten gotree attempts hold zero agree2 risk across every checkpoint, one of them through all 24 checkpoints of a half-successful attempt, while sibling runs have non-zero readings.

\rqanswer{RQ1}{Scanner findings coexist with functional success. Agree2 is positive in 32/830 SWE-bench runs and 21/80 mostly working ProgramBench runs. Both are endpoint measurements. The latter count falls to 5/80 when three advisory-heavy classes are excluded. The benchmarks do not support a controlled comparison of task horizons.}

\rqanswer{RQ2}{Same-task SWE-bench runs differ in endpoint scores, including eight non-zero runs and one zero run on hatch-422. This motivates investigating lower-scoring alternatives. It does not show that agents can reliably be steered toward them, that findings are agent-introduced, or that different ProgramBench runs have identical functionality.}

\section{A Long-Horizon Case Study}
\label{sec:casestudy}

The scale results describe endpoint scanner prevalence. Recorded intermediate states reveal when findings appear and whether they persist. ProgramBench does not publish intermediate code checkpoints, but its public trajectories record every shell command the agent ran, and that turns out to be enough. In the run we study here the agent creates its implementation file with a single heredoc write and afterwards modifies it only through small self-contained text edits and \texttt{sed} one-liners. Replaying those write and edit operations in a sandbox reconstructs the exact state of the file after every agent action, indexed in execution order, and we validate the reconstruction by checking that the replayed final file is byte for byte identical to the published final workspace. For this run, reconstruction adds a per-action risk curve to the endpoint score. Following the original analysis, we carry the last valid reading through temporary syntax errors, so those states are not independent measurements. The official pass ratio is still available only at the end.

\textbf{The run.} We picked the highest-debt run in the working tier, the GPT-5.5 xhigh reimplementation of \texttt{gowsdl}, a tool that fetches a WSDL service description, possibly over the network, and generates Go client code from it. The run passes 90.5\% of the official hidden tests and banks an agree2 SDLI of 10.85 and a union SDLI of 12.66, the maximum in Table~\ref{tab:agg_results} under both channels. The task is security-relevant by nature. The program downloads remote XML and parses it, so transport security and XML handling merit review.

\begin{figure*}[t]
\centering
\includegraphics[width=\textwidth]{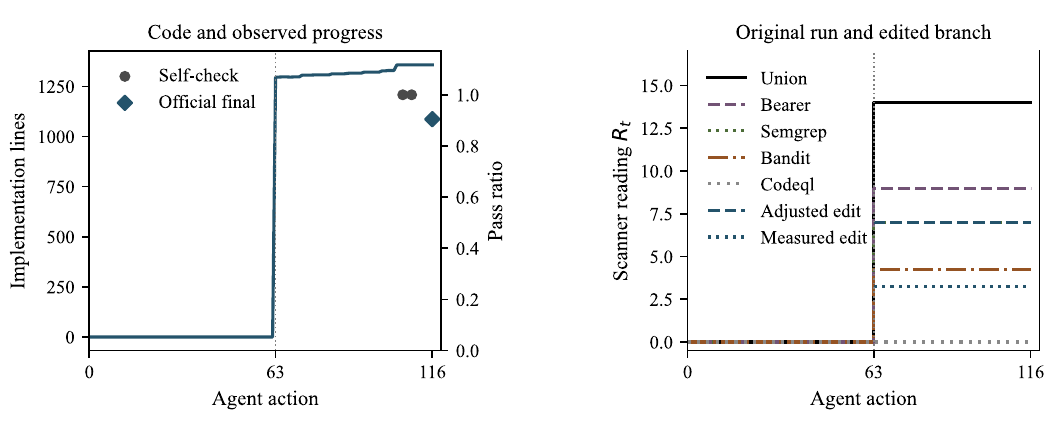}
\caption{Reconstructed gowsdl run. Action 63 writes the implementation, and the recorded readings then remain constant, with the last valid reading carried through syntax errors. Self-check pass ratios are separate from the official final evaluation. The edited branch shows both the measured union score (3.25) and the adjusted score (7.0) retaining three advisory classes. Neither score establishes exploitability.}
\label{fig:gowsdl_case}
\end{figure*}

\textbf{Sixty-three actions of caution, one action of debt.} Figure~\ref{fig:gowsdl_case} shows the anatomy. For its first 62 actions the agent writes no implementation code at all. It probes the reference executable, catalogs flags, renders sample services, and diffs outputs. At action 63 it writes the entire 1{,}296-line implementation in one shot, and that single write already carries every CWE class any scanner ever reports on this run. The reports concern a TLS verification bypass behind the tool's \texttt{-i} flag (CWE-295), \texttt{xml.etree} parsing (CWE-611, CWE-20), URL and file-path handling (CWE-22), and a \texttt{gofmt} subprocess call (CWE-77). These rule labels do not establish exploitability. The remaining 53 actions are pure functional refinement. The agent nudges type naming, documentation placement, and edge cases in small verified steps until its own comparison harness passes all 11 bundled services, and the official evaluation lands at 90.5\%. Three of the four scanners report overlapping CWE classes. Bandit reports all five classes at medium severity and below, Bearer also flags the TLS bypass and rates the path handling and the subprocess call high, Semgrep independently flags the TLS bypass and the XML parsing at high severity and adds a dynamic-URL audit signal (CWE-939), while CodeQL reports nothing on this file. Across all of those actions the risk reading never moves. The union $R_t$ steps from 0 to 14.0 at action 63, the agree2 reading to 12.0, and both stay there to the end of the run.

\textbf{Fidelity pressure and the scanner signal.} The most telling finding is the TLS bypass. The original gowsdl exposes a \texttt{-i} flag that skips certificate verification, so the agent faithfully recreates it using \texttt{ssl.\_create\_unverified\_context}, and three independent scanners flag exactly that idiom. The XML parsing debt has the same character, since the reference tool's behavior is easiest to clone with the permissive standard library parser. Preserving the reference interface can therefore conflict with removing a scanner finding. Whether the optional bypass is a vulnerability depends on the intended trust boundary and deployment.

\textbf{Persistence in the available MirrorCode traces.} The available MirrorCode checkpoints also show persistent readings. Across all 463 scanned checkpoints in its 13 public runs, the agree2 risk reading never decreases even once, six of the seven agree2-debt-carrying runs already show the agree2 pattern at the first scanned checkpoint, and the latest entry we observe anywhere is checkpoint 23 of 80. The union reading decreases exactly once in 463 checkpoints, when a pkl refactor deleted flagged path-handling code along with its advisories. None of the 35 recorded context compaction events coincided with a decrease in agree2, and one tracked HIGH-severity weak-crypto pattern grew from 2 to 6 call sites by run end. These traces show persistent class-level readings. They do not establish that all findings persist at the same sites or that sudden entry is universal.

\rqanswer{RQ3}{In the gowsdl case, all observed classes first appear with one implementation write and persist to the final state. The agree2 reading never decreases across the available MirrorCode checkpoints. This supports inspecting early implementation states, while sparse sampling and class-level aggregation limit conclusions about individual findings.}

\section{Is the Debt Repairable Once Pointed Out?}
\label{sec:repair}

RQ3 says the entry step is observable. RQ4 asks what happens if we act on it. We rewind the gowsdl trajectory to action 63, the first action with scanner findings, and hand the freshly written file to a repair model together with security feedback. The original agent is not available to us, so Gemini 3.1 Pro plays the repairer, and the repair is delivered as exact search and replace edit blocks, the way coding agents actually edit files. Suppressing the scanner is forbidden and does not help, since we strip \texttt{nosec} comments before re-scanning. We compare four feedback arms. The exact findings with rule, CWE, line, and message, or only the sentence that a scanner flagged the file, each as a single attempt and as a retry loop of up to five rounds. Every candidate is verified four ways, a fresh scan, a compile check, byte-level comparison of the generated Go output on all 11 bundled services, and a replay of the agent's 18 later edits on top of the fix. Table~\ref{tab:repair_arms} summarizes the arms. Full prompts and candidate files are in the artifact.

\begin{table}[t]
\centering
\small
\begin{tabular}{>{\raggedright\arraybackslash}p{0.27\columnwidth}>{\raggedright\arraybackslash}p{0.62\columnwidth}}
\toprule
Feedback and budget & Observed outcome \\
\midrule
Precise, one attempt & Parser and URL changes, with a rewritten TLS context. Local output preserved. \\
Precise, up to five & Same checks passed in one round with an independently generated candidate. \\
Vague, one attempt & Parser change only. TLS bypass remained flagged. \\
Vague, up to five & Additional TLS changes by round two. Advisory classes remained. \\
\bottomrule
\end{tabular}
\caption{Repair arms on one gowsdl state using Gemini 3.1 Pro. The prompt required preserving the \texttt{-i} option and byte-identical local WSDL output, and forbade scanner suppression.}
\label{tab:repair_arms}
\end{table}

\textbf{Precise feedback reduces the scanner signal.} With exact findings, the single-attempt arm changed the parser to \texttt{defusedxml}, added an http/https URL-scheme allowlist, and replaced the TLS context construction. It retained certificate-verification bypass under \texttt{-i}, as the prompt required. Thus the changed TLS idiom should not be described as removing that capability. The generated Go output remained byte-identical on 11 local services, and all 18 later edits replayed successfully. These checks establish local behavioral compatibility, not security against adversarial inputs or equivalent hidden-test performance.

The vague single-attempt arm changed only the XML parser. The vague retry arm made additional TLS changes on its second round. This case suggests that feedback detail can affect a repair attempt, but one run and one repair model cannot establish general success rates or the advantage of intervening early.

\textbf{The scanner can be satisfied without being right.} In this case, 16 of the 21 initial findings belong to the advisory-heavy classes CWE-22, CWE-77, and CWE-939. These reports flag operations that may be legitimate for a CLI tool and need contextual review. We do not treat every finding in these classes as harmless, or assume behavior-preserving repair is impossible.

The precise-feedback candidate routed fetching through \texttt{build\_opener().open()}, after which some Bandit and Semgrep matches disappeared. The measured union reading fell to 3.25 even though the relevant fetching capability remained. Bearer still flagged path handling. For the illustrative repaired branch in Figure~\ref{fig:gowsdl_case}, we retain the three advisory classes at their original weights, yielding 7.0. This adjusted reading is an explicit accounting choice, not a measured post-repair scanner score or a lower bound on true risk. Multiple tools exposed this mismatch, but tool agreement alone cannot validate a repair.

\rqanswer{RQ4}{This single case demonstrates edits that reduce scanner findings while preserving tested local behavior. It also shows that an equivalent API rewrite can lower a score without removing a security-relevant capability. Exploit validation and broader repair trials are needed before claiming reliable vulnerability repair.}

\section{Implications for Trustworthy Agentic Systems}
\label{sec:discussion}

\subsection{What Checkpoint Monitoring Adds}
The gowsdl replay locates persistent findings at action 63, before the remaining 53 actions of functional refinement. A final-state scan would detect these same classes. The demonstrated benefit is earlier visibility and localization to an implementation action, not additional detection coverage. Our large-population scores are themselves endpoint readings and cannot establish an advantage over final-only scanning. We have not quantified transient findings missed at the endpoint, compared early and late repair cost, or tested whether earlier feedback improves repair success.

\subsection{Adoption Paths for SDLI}

We see three possible uses for SDLI, each requiring further evaluation.

\textbf{In-line risk gauge in agentic IDEs.} SDLI can be surfaced alongside the existing test-pass gauge in tools like Cursor, Windsurf, and Copilot Workspace. The lightweight Bandit, Bearer, and Semgrep scans at each checkpoint complete in seconds, fast enough for real-time feedback, and agree2 excludes classes reported by only one tool. When $R_t$ rises, the IDE can warn the developer before the agent commits further to a risky architectural path.

\textbf{Stop-condition for autonomous agent loops.} A monitoring policy could pause an agent when current risk rises or cumulative SDLI exceeds a budget. Current risk must be checked separately because SDLI ignores increases after progress plateaus. Our case study identifies a possible intervention point, without establishing optimal timing. This is especially relevant for long-running autonomous agents where human oversight is deferred~\citep{takerngsaksiri2025human}.

\textbf{Reward-shaping signal for agent post-training.} The same-task variation in Table~\ref{tab:swe_instances} motivates testing whether validated lower-risk alternatives can inform preference-based training. SDLI as a per-step penalty ($R_t \cdot \Delta G_t$) can shape agent policy through reinforcement or preference-based learning~\citep{schulman2017proximal,rafailov2023direct,hong2024orpo}, but we have not tested such training. Rewards based only on scanner readings may favor detector evasion.

\subsection{Future Work}
\label{sec:futurework}

Broader intervention studies should cover multiple tasks and repair models, building on self-refinement work~\citep{shinn2023reflexion,madaan2023self,chen2023teaching,he2023large}. Replaying additional ProgramBench runs would provide risk traces, but full SDLI also requires reliable intermediate progress. Other priorities are extending language coverage and evaluating human review of advisory findings at their point of entry.

\subsection{Limitations}
\textbf{Validity and attribution.} We have not performed a stratified manual validation or exploit-based assessment of the population findings. Two tools may report different sites within one class, and our grouping ignores multiplicity. The reported rates therefore concern SAST findings. We did not fingerprint findings against initial repositories or classify them as pre-existing, introduced, resolved, or reintroduced. SWE-bench findings may predate the agent, and transcript extraction can omit code needed to assess a flow. These limitations prevent attributing the measured risk to agent changes.

\textbf{Coverage and comparability.} The oracle covers Python, and its output depends on rules, weights, and scan scope. Advisory-heavy classes affect both union and agree2. The large populations use endpoint scores, while MirrorCode has heterogeneous checkpoints and recorded progress proxies. Missing intermediate states can hide both entry and removal. We have not evaluated checkpoint-frequency sensitivity or total monitoring overhead. Repeated runs share tasks and models, so the rates describe these selected artifacts rather than independent samples of deployments. The three-run MirrorCode working subset is too small for a stable population estimate.

\textbf{Intervention evidence.} The gowsdl run was selected for its high score. Its repair uses one model, with compatibility checked on 11 bundled services rather than the official hidden tests. Retaining \texttt{-i} preserves a TLS bypass, and fewer scanner matches do not prove fewer exploitable defects. The experiment demonstrates feasibility of localized feedback and edits, without establishing reliable steering, general repair success, or a cost advantage over endpoint repair.

\section{Conclusion}
\label{sec:conclusion}

We introduced SDLI to couple scanner readings with increases in historical-best functional progress. Endpoint measurements show that working solutions can carry SAST findings, while MirrorCode and a reconstructed ProgramBench case expose their persistence over recorded states. Same-task variation motivates studying lower-risk alternatives, but does not establish steerability. The repair case preserves tested local behavior while reducing scanner readings, and also exposes how equivalent code can evade pattern matching. Validating findings, attributing them to changes, and comparing checkpoint intervention with final-state scanning remain necessary steps toward a security metric with stronger operational meaning.

\section*{Reproducibility}
SWE-Agent trajectories,\footnote{\url{https://huggingface.co/datasets/nebius/SWE-agent-trajectories}} MirrorCode,\footnote{\url{https://epoch.ai/blog/mirrorcode-preliminary-results}} and ProgramBench submissions\footnote{\url{https://github.com/ProgramBench/submissions}} are public. Our SDLI scripts, per-trajectory BBCS scans, the case study replay, the repair experiment with full prompts and every candidate file, and all intermediate artifacts are archived on Zenodo.\footnote{\url{https://doi.org/10.5281/zenodo.19886200}} The configurations specify Bandit 1.8.0, Bearer 1.49.0, and CodeQL 2.20.0. Semgrep is recorded as 1.101.0 in the SWE-bench/MirrorCode configuration and 1.170.0 in the ProgramBench configuration.

\bibliography{references}

\end{document}